\documentclass[%
 aip,
 amsmath,amssymb,
reprint, 
]{revtex4-2}

\makeatletter
\AtBeginDocument{\let\LS@rot\@undefined}
\makeatother

\usepackage[margin=1in]{geometry}
\usepackage[pdftex]{graphicx}
\usepackage{lipsum}
\usepackage{array}
\usepackage[margin=1in]{geometry}
\usepackage{amsfonts,amsmath,amssymb}
\usepackage[none]{hyphenat}
\usepackage{fancyhdr}
\usepackage{enumitem}
\usepackage{hyperref}
\usepackage{csquotes}
\usepackage{steinmetz}
\usepackage{pdfpages}
\usepackage{xcolor}
\usepackage{soul}
\usepackage{color}
\usepackage{amsmath}
\usepackage{amssymb}
\usepackage{amsfonts}
\usepackage{tikz,array,calc}
\usepackage{arydshln}
\usepackage{cancel}
\usepackage{mathtools}

\usepackage{mathtools}
\usepackage{cancel}

\newcolumntype{C}[1]{>{\centering\let\newline\\\arraybackslash\hspace{0pt}}m{#1}}
\usepackage[T1]{fontenc}
\usepackage{graphicx}
\usepackage{fullpage}
\usepackage{amsmath}
\usepackage{enumitem}
\usepackage{upquote}
\usepackage{float}
\usepackage{caption}
\usepackage{comment}
\usepackage{amsmath}

\def \bF {\pmb{F}}

\def \bx {\pmb{x}}
\def \br {\pmb{r}}
\def \bg {\pmb{g}}
\def \bu {\pmb{u}}
\def \br {\pmb{r}}
\def \bw {\pmb{w}}
\def \by {\pmb{y}}

\def \std {\textnormal{std}}

\usepackage{algorithm}
\usepackage{algpseudocode}

\usepackage{varwidth}

\begin{document}

\title{Synchronizing Chaos using Reservoir Computing}
\author{Amirhossein Nazerian}
	\affiliation{Mechanical Engineering Department, University of New Mexico, Albuquerque, NM, 87131}
\author{Chad Nathe}
	\affiliation{Mechanical Engineering Department, University of New Mexico, Albuquerque, NM, 87131}
\author{Joseph D. Hart}\affiliation{US Naval Research Laboratory, Washington, DC 20375}
\author{Francesco Sorrentino}
	\affiliation{Mechanical Engineering Department, University of New Mexico, Albuquerque, NM, 87131}

\begin{abstract}
 We attempt to achieve isochronal synchronization between a drive system unidirectionally coupled to a response system, 
 under the assumption that limited knowledge on the states of the drive is available at the response.     
    Machine learning techniques have  been previously implemented to estimate the states of a dynamical system from limited measurements. We consider situations in which knowledge of the non-measurable states of the drive system is needed in order for the response system to synchronize with the drive. We use a reservoir computer to estimate the non-measurable states of the drive system from its measured states and then employ
    these measured states to synchronize the response system with the drive.
\end{abstract}

\maketitle


\textbf{A large literature has investigated synchronization of chaos, see e.g., \cite{aranson1989nontrivial,pikovskii1984synchronization,FujiYama83,afraimovich1986stochastic,pecora1990synchronization,pecora1998master,nazerian2022matryoshka} and the review paper \cite{pecora2015synchronization}.
However, little attention has been devoted to the usage of machine learning techniques to enable synchronization. A typical problem is that of synchronizing two identical chaotic systems, 
a drive system that is unidirectionally coupled to a response system. This is typically achieved by having the drive communicate part of its state to the response. Here we use a reservoir observer within a control loop to reconstruct other states of the drive at the response system and use these reconstructed states to aid synchronization. We also show how this proposed scheme can be used to control the time evolution of the response system on an unstable periodic orbit embedded in the chaotic attractor.  The robustness of our proposed approach is studied against measurement noise affecting the information transmitted from the drive to the response system.}

\section{Introduction}

The need of regulating the behavior of nonlinear systems is a common requirement in many physical, social, and biological applications 
\cite{iglesias2010control,del2016control,liu2016control,klickstein2017energy}.
Observers are  classically used in controls applications to estimate states of a dynamical system that cannot be directly measured; feedback control can then be performed using the states that are directly available and those that are reconstructed using the observer. An important problem in the control of nonlinear systems, then, is the construction of a sufficiently accurate observer.

In the absence of a physical model of the system to be observed, one must turn to a data-driven approach. One such approach that is particularly effective when limited data is available is that of reservoir computing. A reservoir computer is a type of recurrent neural network that is designed to be easy to train \cite{jaeger2001echo,maass2002real}. While reservoir computers are most often used in an autonomous mode \cite{pathak2017using}, they have also been found to be an efficient and effective observer of dynamical systems such as Lorenz oscillators \cite{lu2017reservoir,carroll2022time,hart2023time}, semiconductor lasers \cite{cunillera2019cross}, ecological models \cite{kong2023reservoir}, and spatiotemporal systems \cite{lu2017reservoir,zimmermann2018observing}. 

{There is ample observation of synchronization in the natural world \cite{SYNCBOOK}, yet it is unclear whether this observed synchronization may be the result of an underlying learning process.} Autonomous reservoir computing models have been found to be capable of synchronizing with the nonlinear systems upon which they are trained  \cite{antonik2018using,weng2019synchronization} as well as with other identical reservoir computing models \cite{hu2022synchronization,weng2023synchronization,hart2023estimating}. A modification of reservoir computing termed deep reservoir computing has been used for direct learning of a control rule \cite{canaday2021model}. Additionally, autonomous reservoir computing has been used for the estimation of unstable periodic orbits of an unknown dynamical system and for the pinning control of a network on to the estimated unstable periodic orbit \cite{zhu2019detecting}.

Despite these successes, and even though the combination of an observer and a controller is a well established paradigm in control theory, little work has been devoted to study how reservoir computing can be used as an observer for control applications. In this work, we train a reservoir computer as an observer, which is used to mediate the pinning control of a response system by a drive system. The reservoir observer, driven by the measured variable of the drive system, is used to estimate unmeasured variables of the drive system. The measured variable and the estimated variables are then used to drive the response system to a controlled trajectory. We find that the introduction of the reservoir computer can lead to a dramatic reduction in the coupling strength required.

More formally, we consider a problem in which a drive system is unidirectionally coupled to a response system. The drive system produces a reference trajectory for the response system with the goal of controlling or synchronizing the time evolution of the response system on that 
of the drive system.  Another reason why this particular setting may be of interest is that by choosing the initial condition of the drive system to lie on an unstable periodic orbit (UPO) it may be possible to control the time evolution of the response to converge on that particular UPO.

 
We take the response and the drive to be described by the same equations and we assume that when these systems are uncoupled, their dynamics evolve on the same chaotic attractor.
We consider the following set of generic dynamical equations that describe the time evolution of the drive, $\bx_D(t)$, and the response, $\bx_R(t)$,
\begin{equation} \label{eq:general}
    \begin{aligned}
\dot{\bx}_D(t) & = \bF(\bx_D(t))\\
\dot{\bx}_R(t) & = \bF(\bx_R(t)) +  \kappa H (\bx_D(t) - \bx_R(t))\\
    \end{aligned}
\end{equation}
where $\bF : \mathbb{R}^m \rightarrow \mathbb{R}^m$ and $m$ is the number of states in the dynamics system. The matrix $H$ is size $m\times m$ and describes the coupling scheme. The scalar $k>0$ is the coupling strength.

In the rest of this paper, without loss of generality we take $m=3$ and write ${\bx}_D(t)=[x_D(t),y_D(t),z_D(t)]$, ${\bx}_R(t)=[x_R(t),y_R(t),z_R(t)]$.
We assume the entries of the matrix $H=\{H_{ij}\}$ to be either zeros or ones, and consider different coupling schemes, as a result of different choices of the matrix $H$.
For example, by setting all the entries of the matrix $H$ equal to zero except for entry $H_{12}=1$, we indicate that the state $y_D(t)$ from the drive system is available to the response system and that $y_D(t)$ appears in the equation of the first state of the response system. We will also refer to this coupling scheme as $y_D(t) \rightarrow x_R(t)$. 
We will instead refer to a coupling scheme as $\hat{y}_D(t) \rightarrow x_R(t)$ when $y_D(t)$ is not a measured variable from the response system, but an estimate of $\hat{y}_D(t)$. 

The rest of this paper is organized as follows. 
In Sec.\,\ref{sec:res}, reservoir computing and the general procedure of the drive-response system are discussed. 
Examples of the Chen and R\"ossler systems are also provided, and the effect of measurement noise on the performance of the reservoir computer is discussed.
Conclusions are provided in Sec.\,\ref{sec:conclusions}.
{In Appendix \ref{appendix}, we discuss how the integration of the continuous time system and the evolution of the discrete time RC are performed.}

\section{Reservoir Computing} \label{sec:res}

We begin by introducing the reservoir equation,
\begin{equation} \label{eq:restr}
\br(t+\Delta t) = (1-\alpha)\br(t) + \alpha \tanh{(A\br(t) + s(t)\bw_{\text{in}})}
\end{equation}
where $A$ is the coupling matrix, $s(t)$ is the drive signal, $\Delta t$ is the discrete time step, and $\bw_{\text{in}}$ is a vector of random elements drawn from a standard Gaussian distribution, i.e., $\mathcal{N}(1,0)$. We set $\Delta t = 0.001$ throughout this paper unless stated otherwise. The matrix $A$ is the adjacency matrix of a directed Erdos Renyi network  with $N$ nodes and connectivity probability $p=0.1$. We make $A$ to be have a spectral radius of $\rho$ via the operation, $A\leftarrow A\rho/ \lambda$, where $\lambda$ is the largest real eigenvalue and $0 < \rho \leq 1$.

From the time evolution of the reservoir equation, Eq.\ \eqref{eq:restr}, one can construct the readout matrix,
\begin{equation}
\Omega=
\begin{bmatrix}
    r_1(1) & r_2(1) & \cdots & r_N(1) & 1\\
    r_1(2) & r_2(2) & \cdots & r_N(2) & 1\\
    \vdots & \vdots & \ddots &  \vdots  & \vdots \\
    r_1(T_1) & r_2(T_1) & \cdots & r_N(T_1) & 1\\
\end{bmatrix}
\end{equation}
where $r_i(t)$ is the readout of node $i$ at time $t$ and $t=T_1$ indicates the end of the training phase. The last column of the matrix $\Omega$ is set to 1 to account for any constant offset in the fit. We then relate the readouts to the training signal, $y(t)$, with additive noise, via the unknown coefficients contained in the vector, $\bw_{\textnormal{out}}$, 
\begin{equation}
    \Omega \bw_{\textnormal{out}} = \mathbf{g} 
\end{equation}
where $\mathbf{g}(t)$ is the training signal. We then compute the unknown coefficients vector $\bw_{\textnormal{out}}$ via the equation,
\begin{equation} \label{kappa}
   \bw_{\textnormal{out}} =  \mathbf{\Omega}^{\dagger}\mathbf{g}.
\end{equation}
Here, $\mathbf{\Omega}^{\dagger}$ is given as,
\begin{equation}
   \mathbf{\Omega}^{\dagger} =  \mathbf{\left(\Omega^T \Omega + \beta I \right)^{-1}\Omega^T}.
\end{equation}
In the above equation, $\beta$ is the ridge-regression parameter used to avoid overfitting \cite{lu2017reservoir}  and $\mathbf{I}$ is the identity matrix. For all the simulations in this work, $\beta=10^{-9}$ is used.
For $\beta=0$, $\mathbf{\Omega}^{\dagger}$ is the pseudo inverse matrix of $\mathbf{\Omega}$. Next, we define the training fit signal as,
\begin{equation} \label{eq:RO}
\mathbf{h} = \Omega \bw_{\textnormal{out}}.
\end{equation} 
Lastly, the training error is computed as,
\begin{equation} \label{tre}
\Delta_{\text{tr}} = \frac{\std(\mathbf{h} - {\mathbf{g}} )}{\std( \mathbf{g})}
\end{equation}
where the notation $\std(\cdot)$ denotes the standard deviation.

The testing phase is carried out in the same manner, except using the previously computed coefficients contained in, $\bw_{\textnormal{out}}$, to compute the fit signal, $\mathbf{h}_{\text{ts}}$ according to the $\Omega_{\text{ts}}$ matrix generated by the new drive signal, i.e., the testing signal $\mathbf{g}_{\text{ts}} (t)$. We call the length of the testing phase $T_s$. 
The testing error is computed as
\begin{equation} \label{eq:ts}
    \Delta_{\text{ts}} = \frac{\std(\mathbf{h}_{\text{ts}} - {\mathbf{g}_{\text{ts}}} )}{\std( \mathbf{g}_{\text{ts}} )}.
\end{equation}

\begin{figure*}
    \centering
    \includegraphics[width=0.7\linewidth]{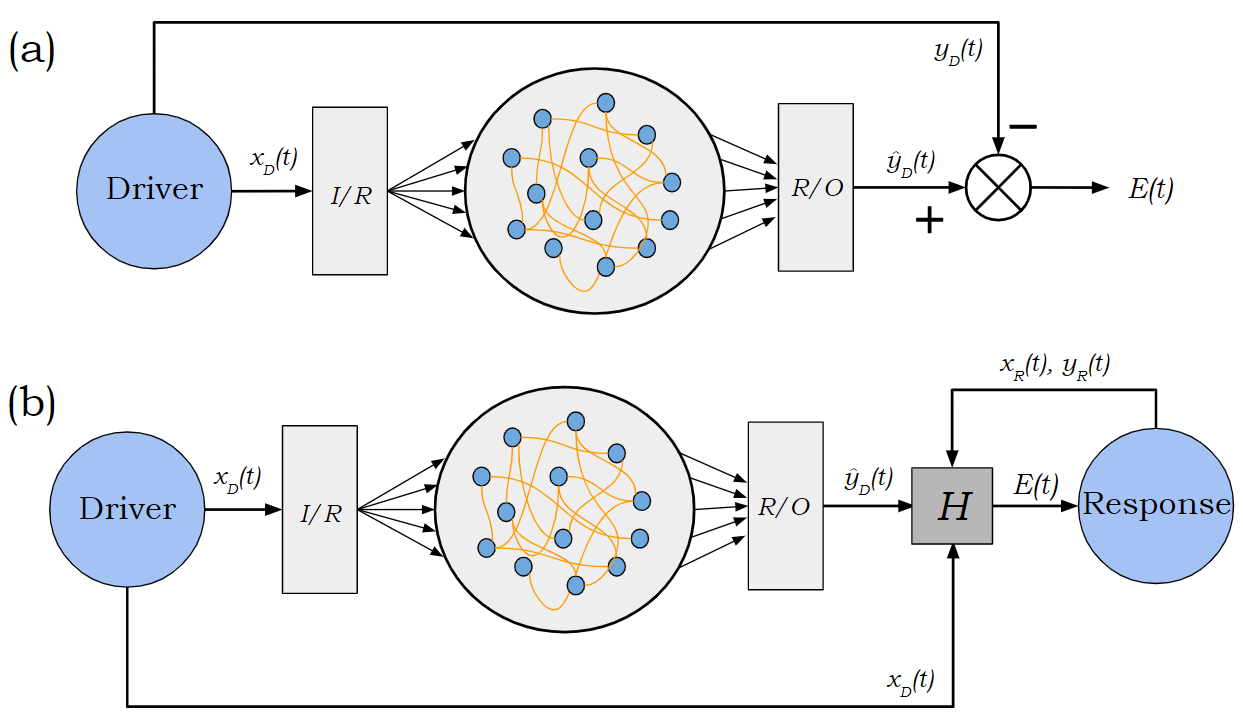} 
    \caption{(a) Training phase of reservoir. The reservoir input signal is $x_D(t)$ and it is trained on $y_D(t)$. The error, $E(t)$, is calculated with the fit signal, $\hat{y}(t)$, and the training signal. (b) Control configuration. The response system takes as an input, $E(t)$, where $E(t)$ is $\bx_R(t) - \bx_D(t)$. The matrix, $H$, describes what state variables are used in coupling.} \label{fig:schematic}
\end{figure*}

In Fig. \ref{fig:schematic}, we show the process of training the reservoir computer in (a) and then using it for control in (b). I/R is the input-to-reservoir function and is represented by the term $s(t)\bw_{\text{in}}$ in Eq. \eqref{eq:restr}. R/O is the reservoir-to-output function represented by Eq. \eqref{eq:RO}.

\subsection{Chen system}
We take both the drive and the response systems to be described by the Chen equation $\dot{\bx} (t) = \bF (\bx(t))$ where $\bx = [x, \ y, \ z]^\top$ and
\begin{equation} \label{eq:Chen}
    \bF (\bx(t)) = 
    \begin{bmatrix}
    a(y(t) - x(t))\\
    (c - a - z(t)) + cy(t)\\
    x(t) y(t) - \beta z(t)
    \end{bmatrix},
\end{equation}
where here we have set $a = 35, c = 28$ and $\beta = 8/3$.
The goal is to synchronize the trajectories of the response system and the drive system, using the scheme shown in Fig.\,\ref{fig:schematic}.
It is assumed that the only component of the drive system accessible to the response system is the time evolution of the state $x_D(t)$.

The master stability function \cite{Pe:Ca} predicts that by coupling $x_D \rightarrow y_R$, the two systems synchronize when the coupling strength $\kappa > 10.62$ \cite{Pecora2009}.
However, if the information about $y_D$ is available, the two systems synchronize by the coupling $y_D \rightarrow y_R$ when $\kappa > 3.54$ \cite{Pecora2009}.
Hence, we would like to estimate $\hat{y}_D$ from $x_D$ in order to synchronize the two systems with a lower coupling strength $3.54 < \kappa < 10.62$ through coupling $\hat{y}_D \rightarrow y_R$. Note that coupling $x_D \rightarrow x_R$ does not result in synchronization for any value of $k$, as shown in \cite{Pecora2009}.


\begin{figure}
    \centering
    \begin{tabular}{l r}
    \text{(A)} \hfill  \text{(B)} \\ 
    \includegraphics[width=.45\linewidth] {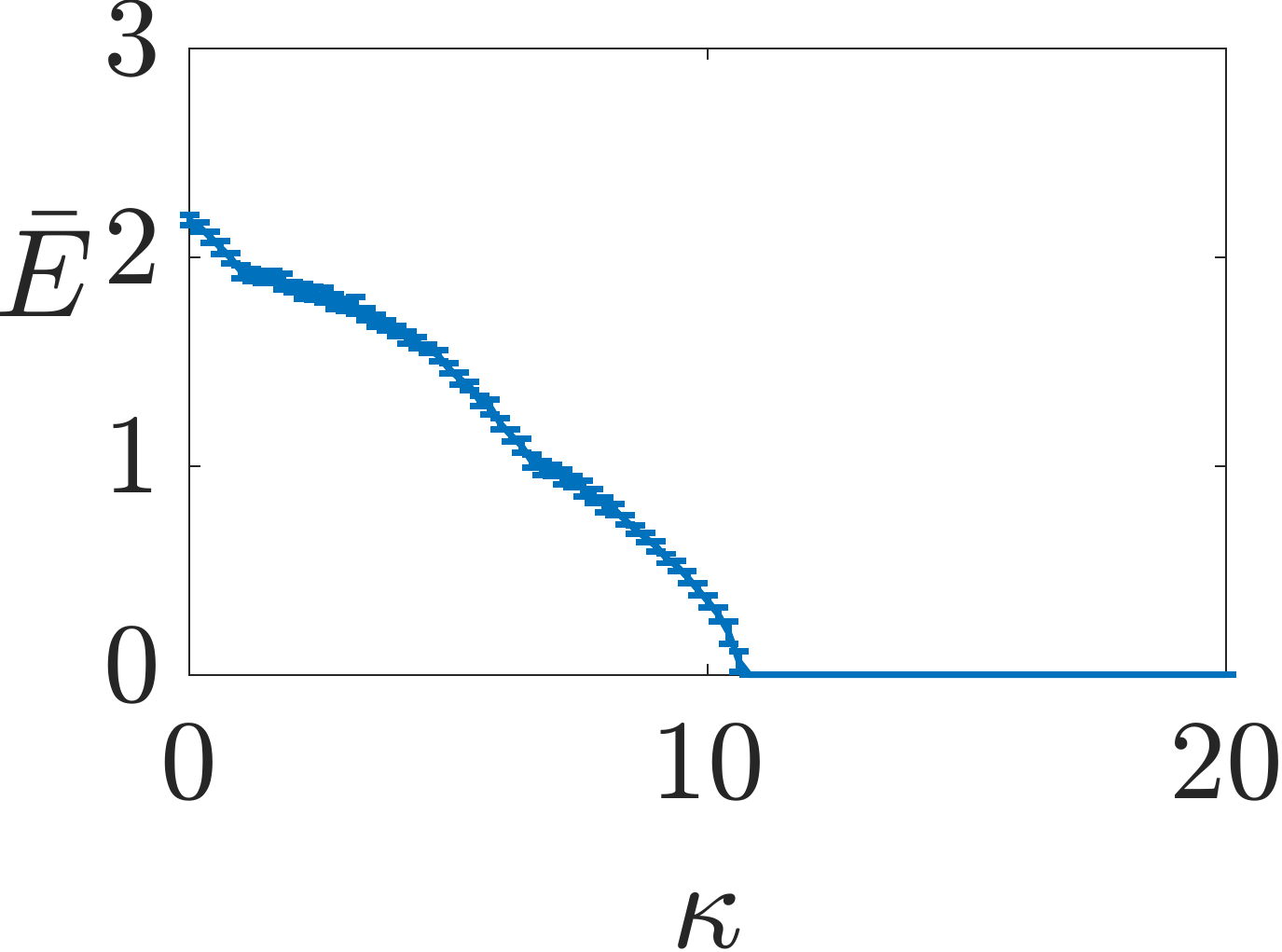} 
    \includegraphics[width=.45\linewidth]{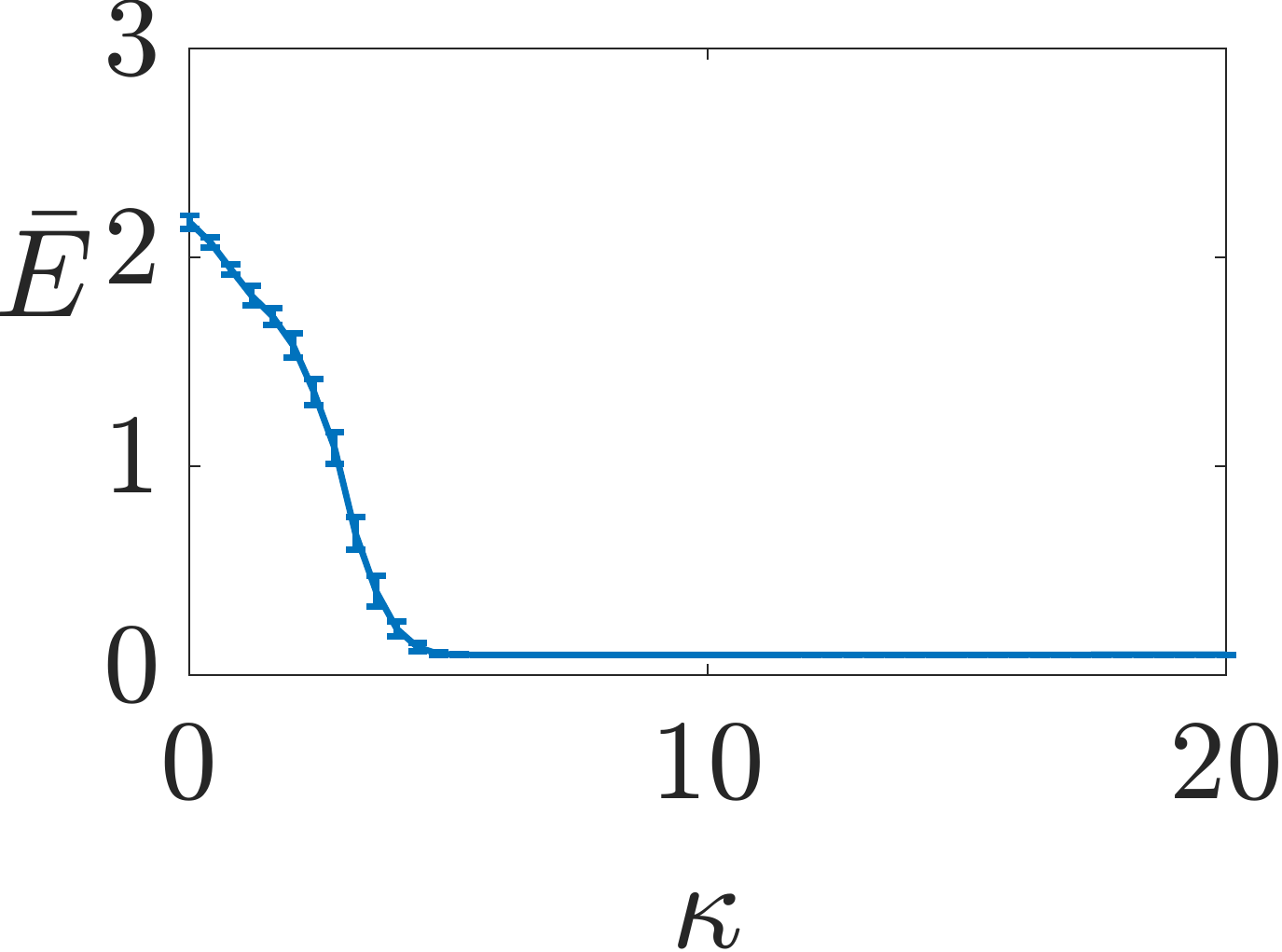} \\  
    \hspace{0.25\linewidth} \text{(C)} \\
    \hspace{0.25\linewidth} \includegraphics[width=.45\linewidth]{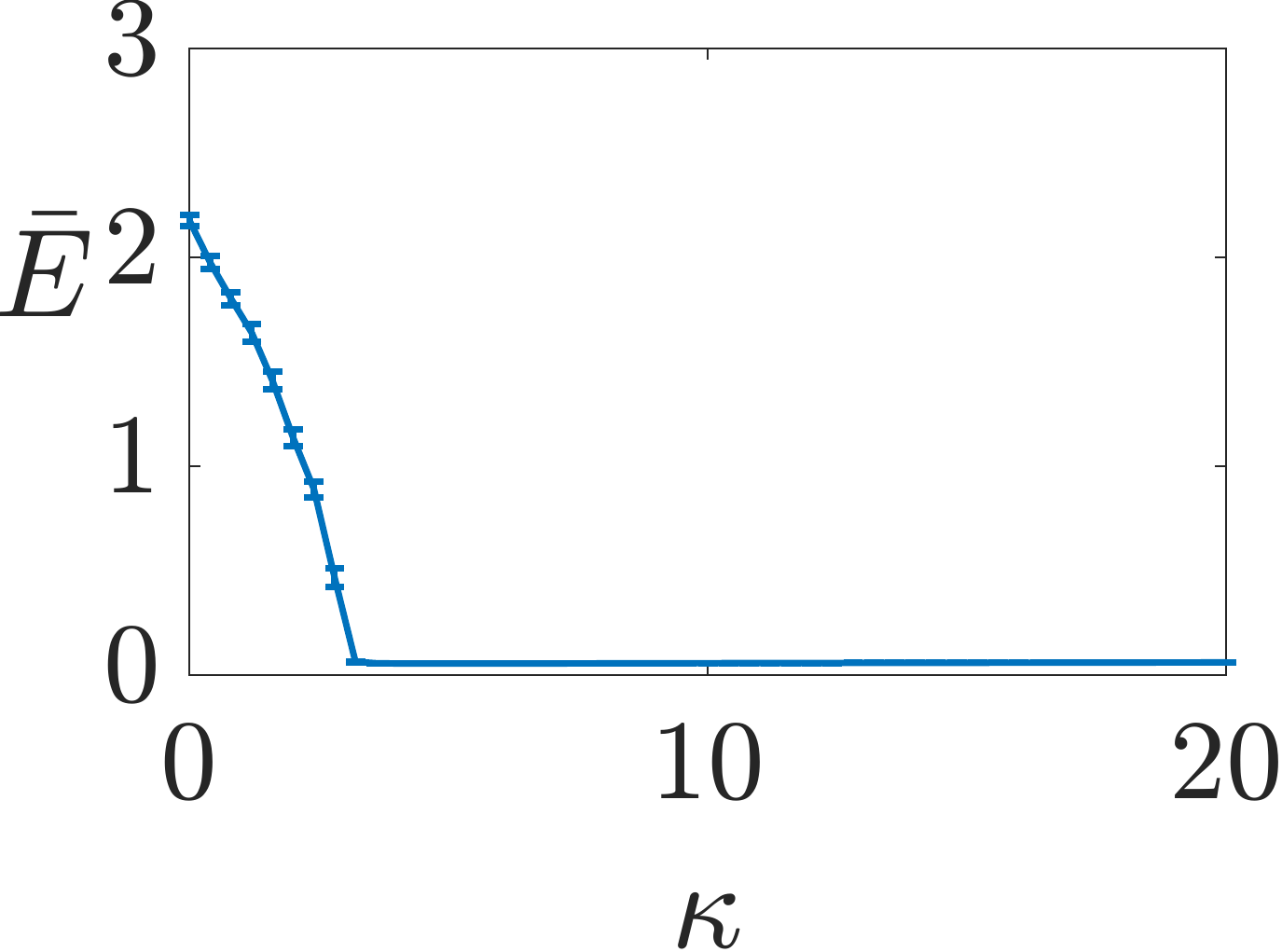}
    \end{tabular}
    \caption{
    Chen system. (A) Coupled $x_D \rightarrow y_R$, $\kappa_{\min} \simeq 10.78$ (B) Coupled $\hat{y}_D \rightarrow y_R, \kappa_{\min} \simeq 3.96$ (C) Coupled both $x_D \rightarrow y_R$ and $\hat{y}_D \rightarrow y_R, \kappa_{\min} \simeq 3.00$. The errorbars represent  standard deviations. } \label{fig:chen}
\end{figure}

We introduce the instantaneous and average synchronization errors,
\begin{align} \label{eq:error}
\begin{split}
    E(t)  := &  \left[ \left( \frac{x_D (t) - x_R(t)}{\std(x_D(t))_t} \right)^2  +  \left(\frac{y_D (t) - y_R(t)}{\std(y_D(t))_t}\right)^2 \right.  \\
    &   + \left. \left(\frac{z_D (t) - z_R(t)}{\std(z_D(t))_t}\right)^2 \right]^\frac{1}{2} \\
\bar{E}  =& <E(t)>_t,    
\end{split}
\end{align}
respectively, where $< \cdot >_t$ indicates an average over the time interval $t = [500, 1000]$ seconds, and $\std(\cdot)_t$ returns the standard deviation (a positive scalar value) of the time-series in the argument over the time interval $t$.

Figure \ref{fig:chen} shows the synchronization error $\bar{E}$ 
for different coupling schemes and different coupling strengths, when the drive and the response system are initialized from randomly chosen initial conditions from the Chen system attractor. The results we obtain are in agreement with the predictions of the master stability function.
From each plot in Fig.\ \ref{fig:chen} we see that there is a critical coupling strength $\kappa_{\min}$ above which a transition from asynchrony to synchrony is observed. Fig.\ \ref{fig:chen} (A) corresponds to coupling $x_D \rightarrow y_R$, for which $\kappa_{\min} \simeq 10.78$; Fig.\ \ref{fig:chen} (B) corresponds to  coupling $y_D \rightarrow y_R$, for which  $\kappa_{\min} \simeq 3.96$. Finally, Fig.\ \ref{fig:chen} (C) corresponds to coupling in both $x_D \rightarrow y_R$ and $y_D \rightarrow y_R,$ for which $ \kappa_{\min} \simeq 3.00$. 

 By comparing Figs.\,\ref{fig:chen} (B) and \,\ref{fig:chen} (C) with Fig.\,\ref{fig:chen} (A) we see that a much lower coupling strength is needed for synchronization when coupling involves the state $y_D(t)$, hence in what follows we attempt to reconstruct 
$\hat{y}_D(t) \approx y_D(t)$ from knowledge of $x_D(t)$.

Our approach illustrated in Fig.\ref{fig:schematic} is to indirectly couple the response system to the drive system through the estimate $\hat{y}_D(t)$ produced by the reservoir observer.
Hence, the coupled system equations are,
\begin{equation} 
    \begin{aligned}
        \dot{\bx}_D (t) & = \bF(\bx_D (t)) \\
        \dot{\bx}_R (t) & = \bF(\bx_R (t)) 
        + \kappa \begin{bmatrix}
        0 \\ \hat{y}_D - y_R + x_D - x_R \\ 0
        \end{bmatrix} \\
    \end{aligned}
\end{equation}
If the two systems are coupled `ideally', then $\hat{y}_D$ is replaced by the true value $y_D$.

\begin{figure}
    \centering
    \includegraphics[width=\linewidth]{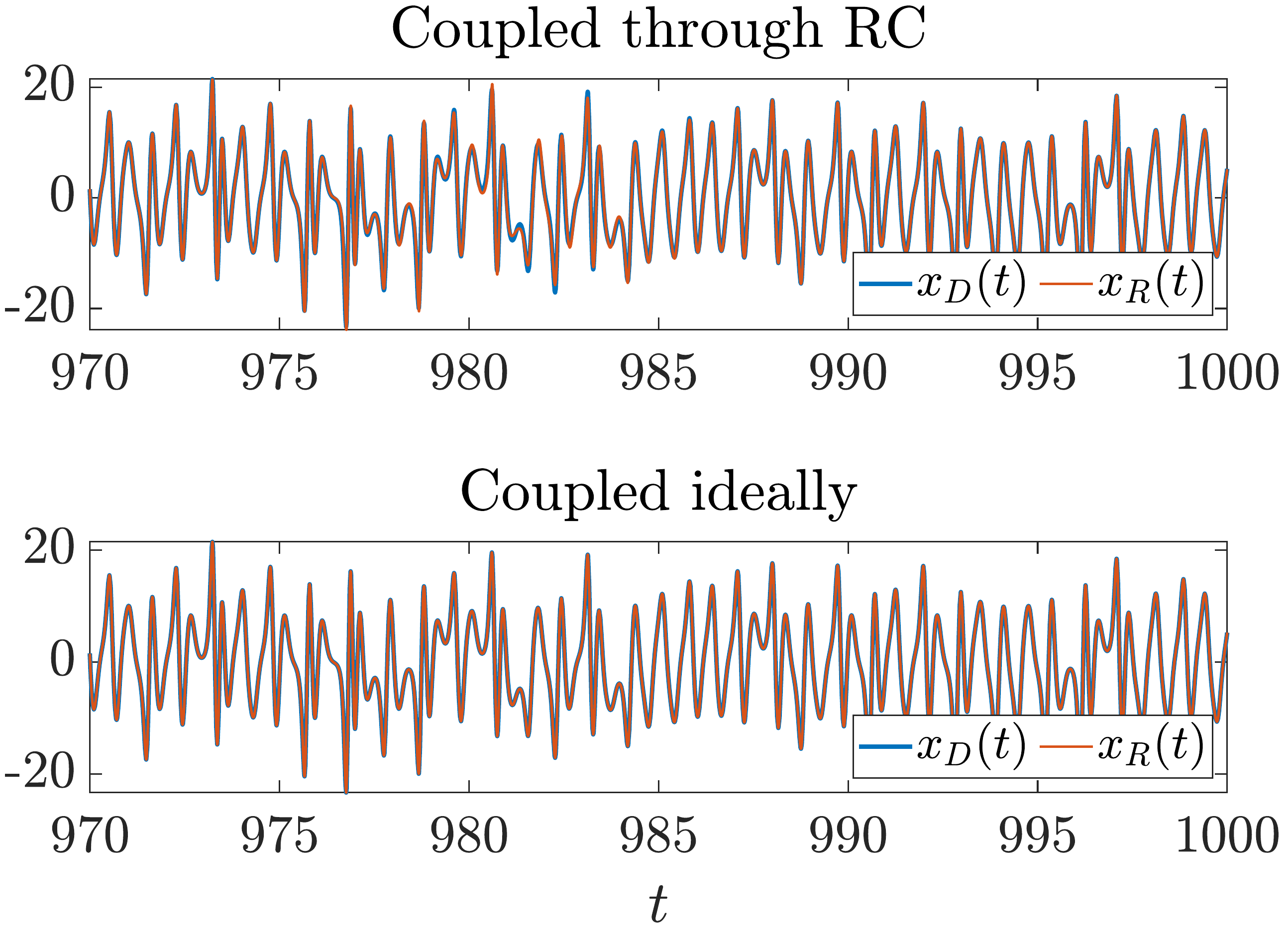}
    \caption{The plots show the time evolution of the $x$ component of the drive and the response systems when coupled through the RC in $x_D \rightarrow y_R$ and $\hat{y}_D \rightarrow y_R$ (top) and when coupled ideally in $x_D \rightarrow y_R$ and $y_D \rightarrow y_R$ (bottom.) 
    Here, the coupling strength $\kappa = 3.1$ for both cases.}
    \label{fig:chenx}
\end{figure}

\begin{figure}
    \centering
    \includegraphics[width=0.9\linewidth]{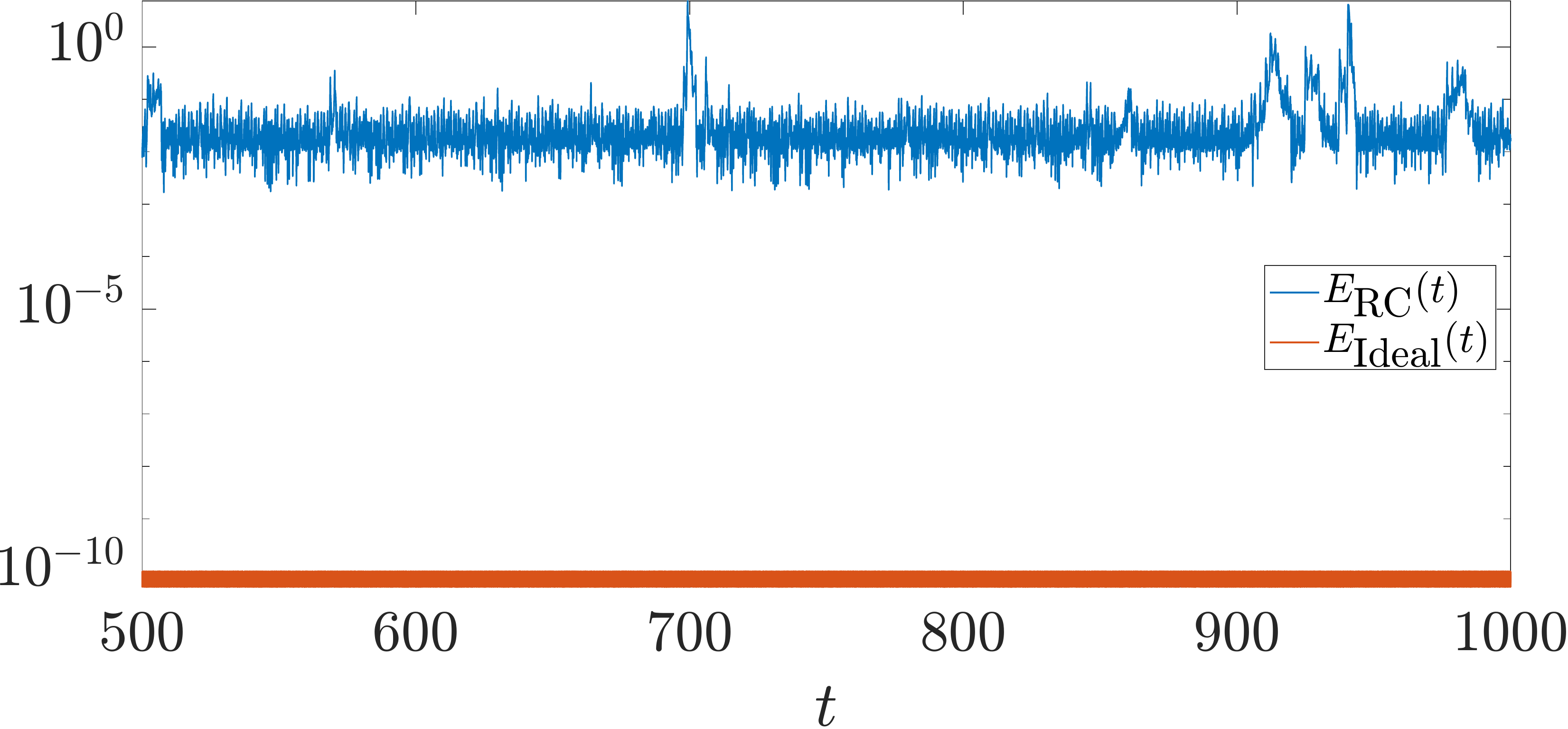}
    \caption{Synchronization error of Chen's drive and response systems from time series in Fig.\,\ref{fig:chenx}. The plot shows $E_{\textnormal{RC}}(t)$, the error when the coupling $x_D \rightarrow y_R$ and $\hat{y}_D \rightarrow y_R$ is through estimation by RC, and $E_{\textnormal{Ideal}}(t)$, the error in the ideal case when all the states of the drive system are known, $x_D \rightarrow y_R$ and $y_D \rightarrow y_R$.}
    \label{fig:chenerror}
\end{figure}

As an example, Fig.\,\ref{fig:chenx} shows the time evolutions of the $x$ components of the drive and response system as they are coupled through the RC and ideally. We tentatively set the RC parameters equal to $\alpha = 0.61$, $\rho = 0.9$, 500 nodes, and the input weights are drawn randomly from a uniform distribution between $-0.5$ and $0.5$. 
The two systems are initialized from different initial conditions on the chaotic attractor of the Chen system.
We see that the two coupling schemes produce comparable levels of synchronization between the drive and response systems, which appear qualitatively similar to the human eye. However, a calculation shows that the two synchronization errors differ by several orders of magnitude. This can be seen 
in 
Figure \ref{fig:chenerror}, which compares the synchronization errors $E_{\textnormal{RC}}(t)$ and $E_{\textnormal{Ideal}}(t)$ through a 1000$s$ simulations. Here, the instantanous synchronization errors are $E_{\textnormal{RC}}(t)$ when coupled through the RC, and $E_{\textnormal{Ideal}}(t)$ when coupled ideally.
We see that while the synchronization error attains a small value throughout the simulation when coupling uses the estimated state, the error attains a much smaller value (equal to the numerical precision of the computer) when coupling uses the true state.
This is due to the fact that the estimate produced by the RC observer is not exactly the same as the true signal, but just an approximation. It is also important to point out that the ideal case shown in Fig.\ \ref{fig:chenerror} does not consider the presence of noise in the signal transmitted from the drive system to the response system. This irrealistic assumption is removed in Sec.\ \ref{sec:noise}.

\begin{figure}
    \centering
    \begin{tabular}{c}
    \text{(A)} \\
    \includegraphics[width=.8\linewidth]{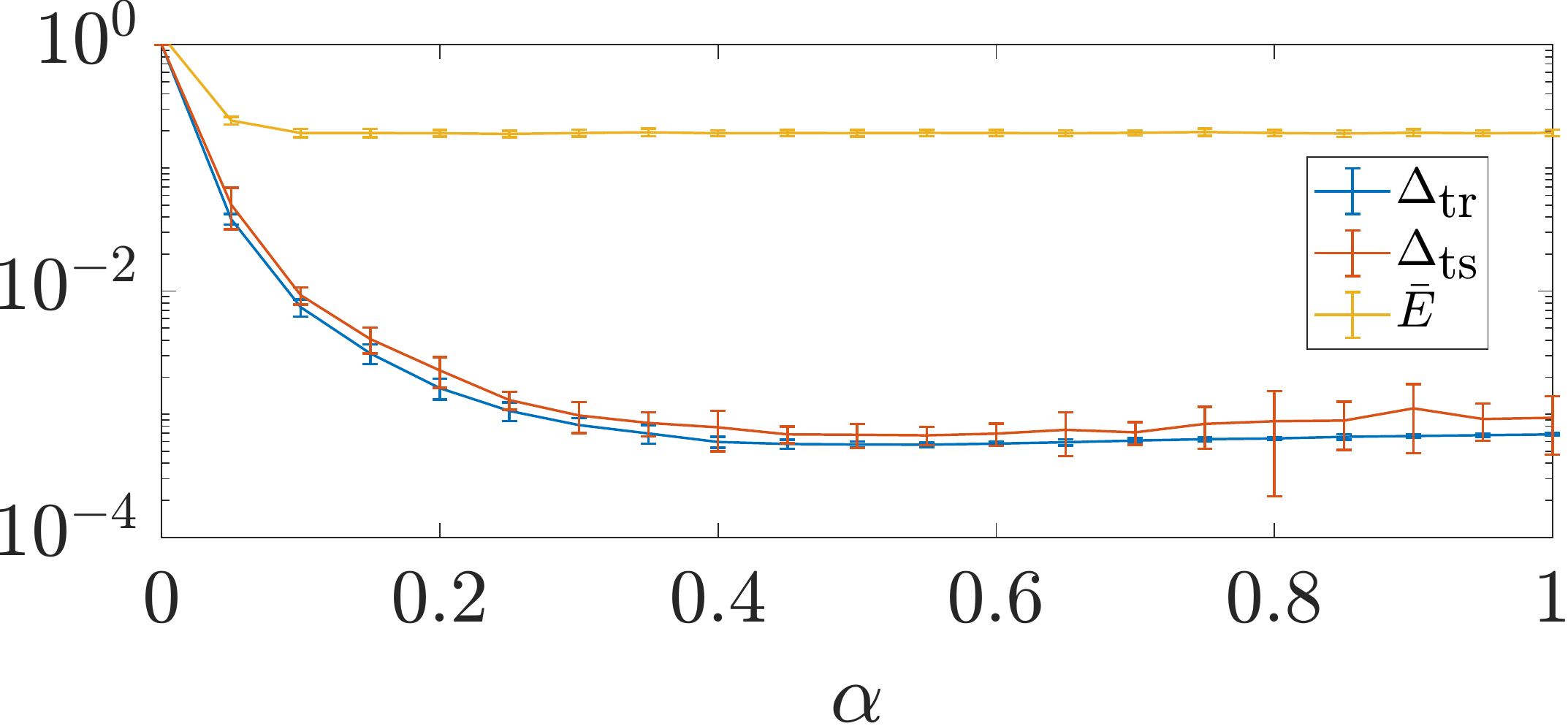}  \\ \text{(B)} \\
    \includegraphics[width=.8\linewidth]{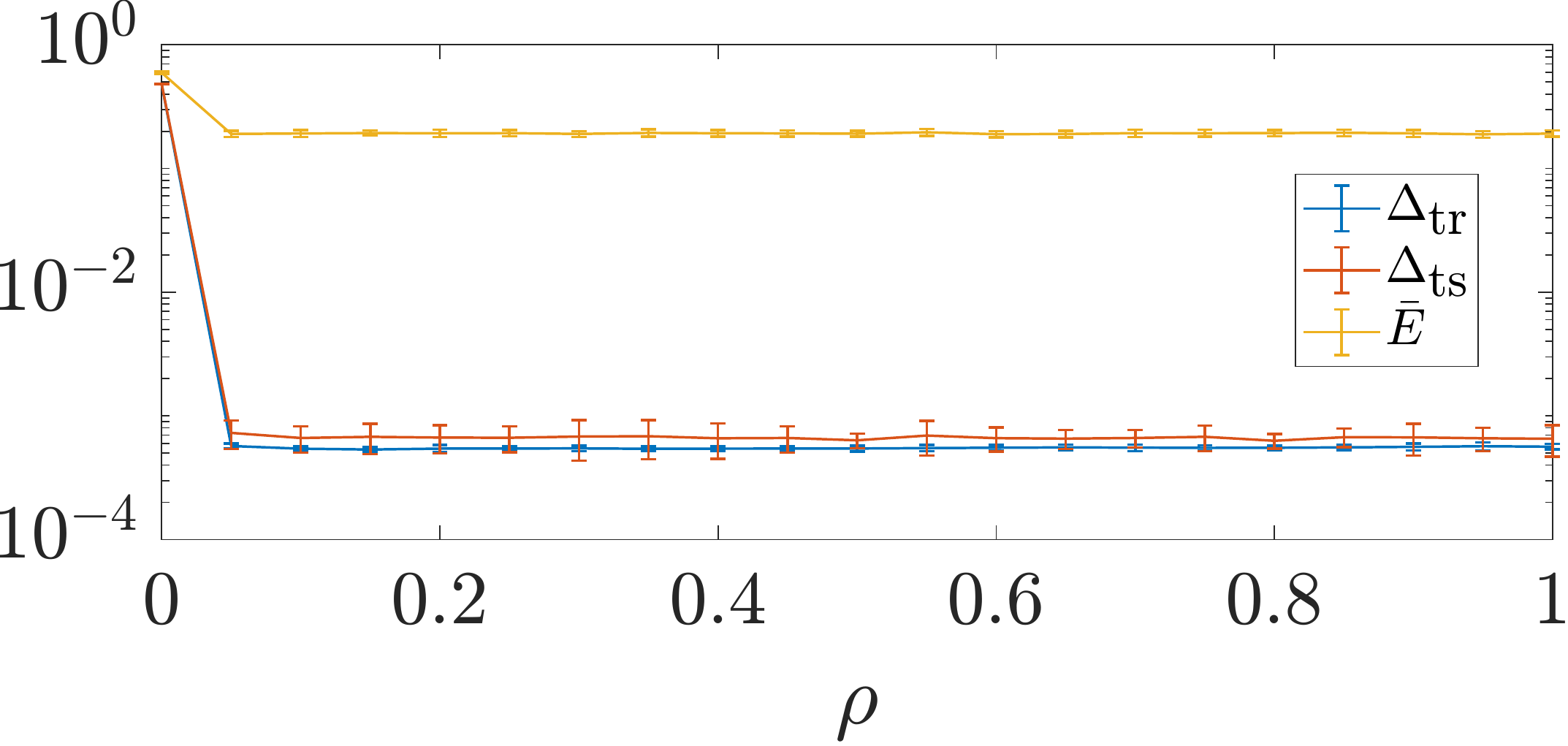} 
    \end{tabular}
    \caption{Chen system. Coupling is in $x_D \rightarrow x_R$ and $\hat{y}_D \rightarrow y_R$ with $\kappa=10$ (A) We plot the training $\Delta_{\textnormal{tr}}$, testing $\Delta_{\textnormal{ts}}$, and synchronization $E$ errors as a function of the leakage parameter, $\alpha$ 
 for a fixed value of the spectral radius $\rho=0.9$. (B)  We plot the training, testing, and synchronization error as a function of the spectral radius, $\rho$, of the $A$ matrix, for a fixed value of the leakage parameter  $\alpha=0.5$.
    } \label{fig:chen_sycnhtrain} %
\end{figure}

The results in Figs.\ \ref{fig:chen} and \ref{fig:chenx} show the feasibility of our approach. An important step is to properly choose the hyperparameters of the RC observer.
Figure \ref{fig:chen_sycnhtrain} (A) shows the training, testing, and synchronization errors, $\Delta_{\textnormal{tr}}$, $\Delta_{\textnormal{ts}}$, and $\bar{E}$, respectively, as the parameter $\alpha$ in \eqref{eq:restr} is varied. 
We see that as $\alpha$ grows to 0.4, the training and synchronization errors are substantially reduced. However, for $0.4 \leq \alpha \leq 1$, the testing errors attain higher standard deviations, especially  for larger values of $\alpha$. 
Figure \ref{fig:chen_sycnhtrain} (B) shows the effect of varying the spectral radius $\rho$ over the training, testing, and synchronization errors, for $\alpha=0.5$. 
We see that the performance we achieve is quite robust to variations in $\rho$, as long as $\rho \geq 0.1$. 
We set $\rho = 0.9$ in our simulations.

\subsection{R\"ossler} \label{sec:ross}

The dynamical equation of a R\"ossler system is $\dot{\bx} (t) = \bF (\bx(t))$ where $\bx = [x, \ y, \ z]^\top$ and
\begin{equation} \label{eq:rossler}
    \bF (\bx(t)) = 
    \begin{bmatrix}
    -y(t) - z(t))\\
   x(t) + a y(t)\\
   b + (x(t) -c)z(t)
    \end{bmatrix}
\end{equation}
where we set $a=b=0.2$ and $c=9$. In what follows, we set the initial conditions of the drive and response system to be randomly chosen points on the chaotic attractor. We examine the effect of the gain, $\kappa$, on the synchronization error in the scenarios where we couple $x_D \rightarrow x_R$, $y_D \rightarrow y_R$, and both simultaneously. We then move on to show the performance of the RC when $x_D \rightarrow x_R$ and $\hat{y}_D \rightarrow y_R$ are the couplings. 

\begin{figure}
    \centering
    \begin{tabular}{c}
    \text{(A)} \\ 
    \includegraphics[width=.8\linewidth]{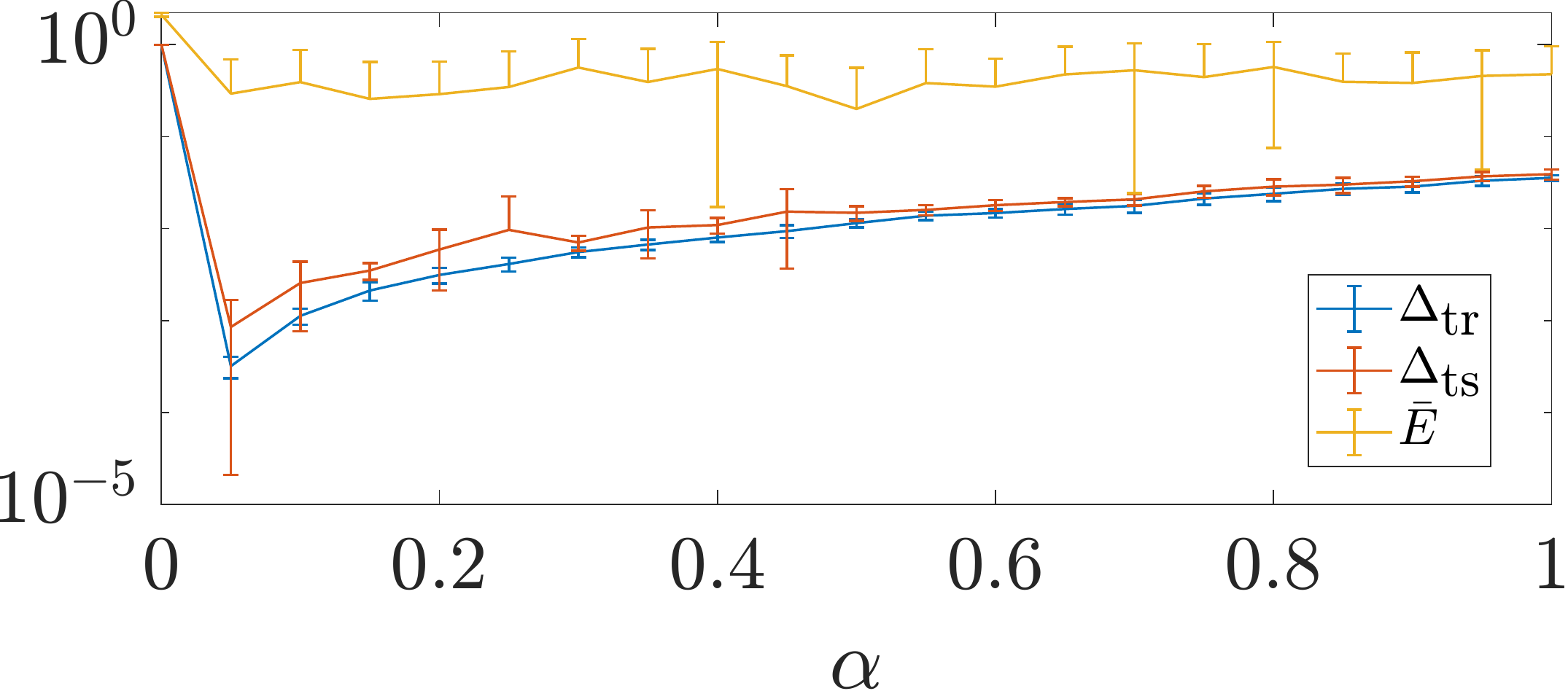}  \\ \text{(B)}\\
    \includegraphics[width=.8\linewidth]{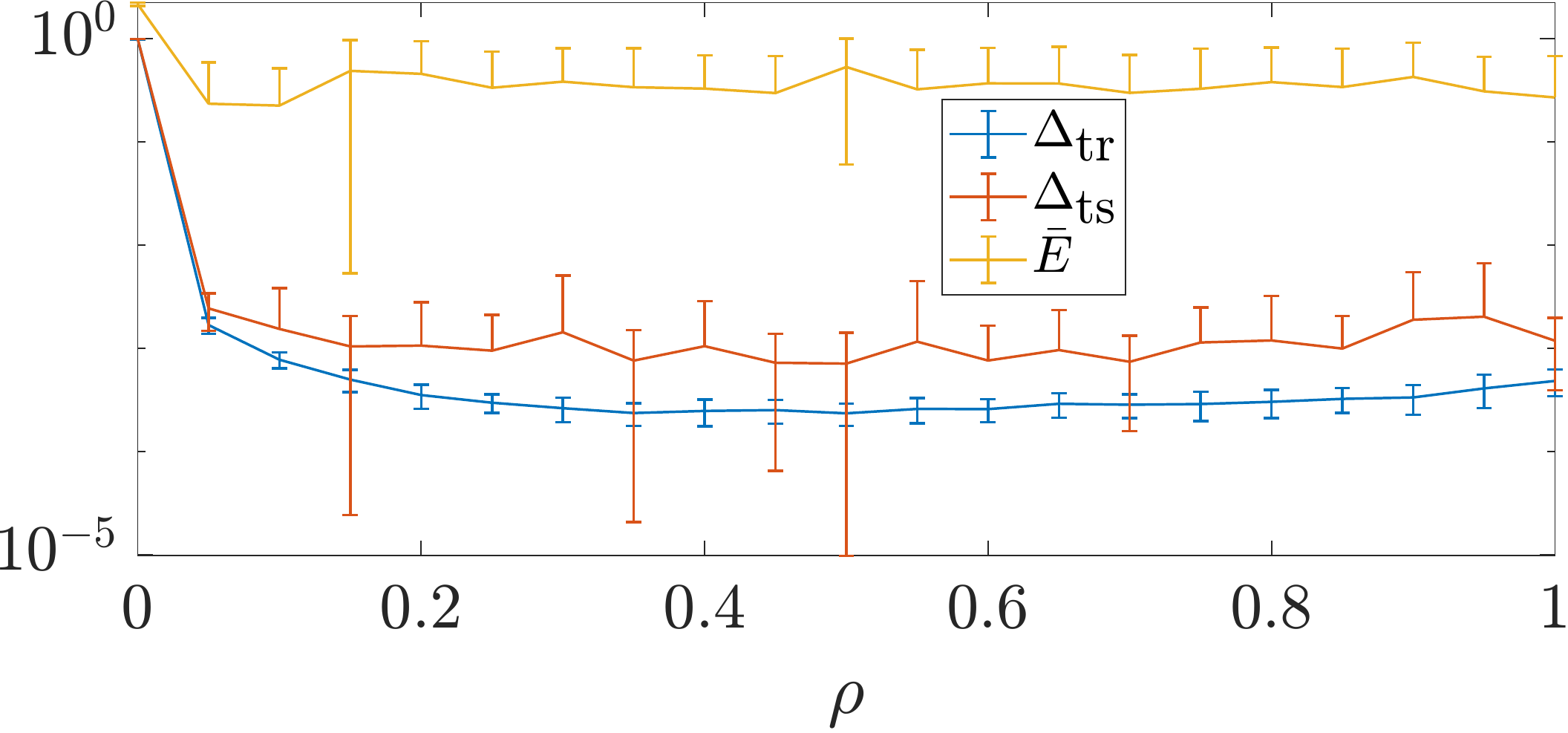} 
    \end{tabular}
    \caption{R\"ossler system. Coupling is in $x_D \rightarrow x_R$ and $\hat{y}_D \rightarrow y_R$ with $\kappa=0.15$. (A) We plot the training, testing, and synchronization error as a function of the leakage parameter, $\alpha$. We set $\rho=0.6$ (B)  We plot the training, testing, and synchronization error as a function of the spectral radius, $\rho$, of the $A$ matrix. We set $\alpha=0.05$} \label{fig:rossleralpharho}
\end{figure}

In Fig.\ \ref{fig:rossleralpharho} we perform a similar study as what previously shown in  Fig.\ \ref{fig:chen_sycnhtrain} to select the hyperparameters of the RC for the case of the R\"ossler system. 

\begin{figure}
    \centering
     \begin{tabular}{l r}
    \text{(A)} \hfill  \text{(B)} \\ 
    \includegraphics[width=.45\linewidth] {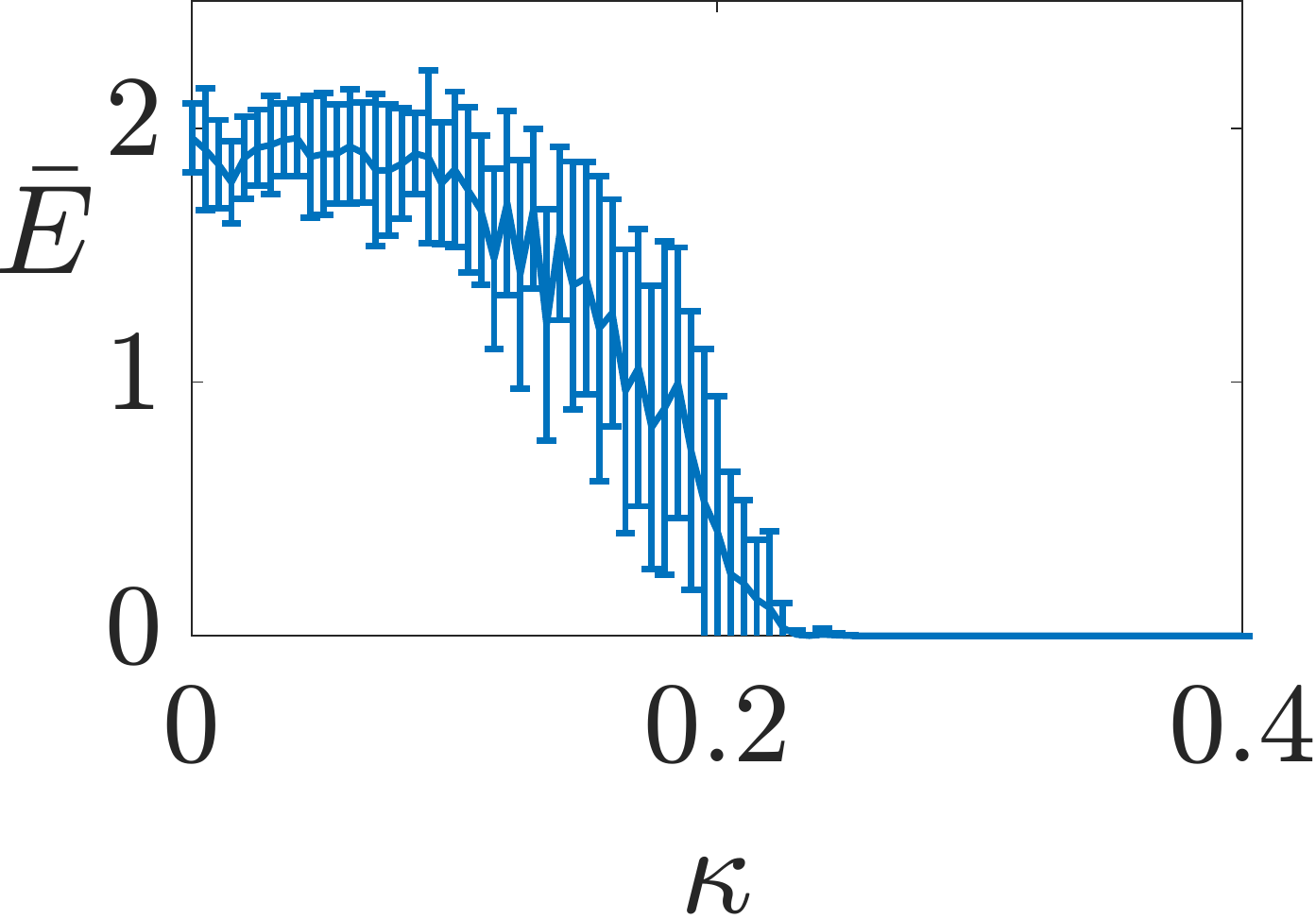} 
    \includegraphics[width=.45\linewidth]{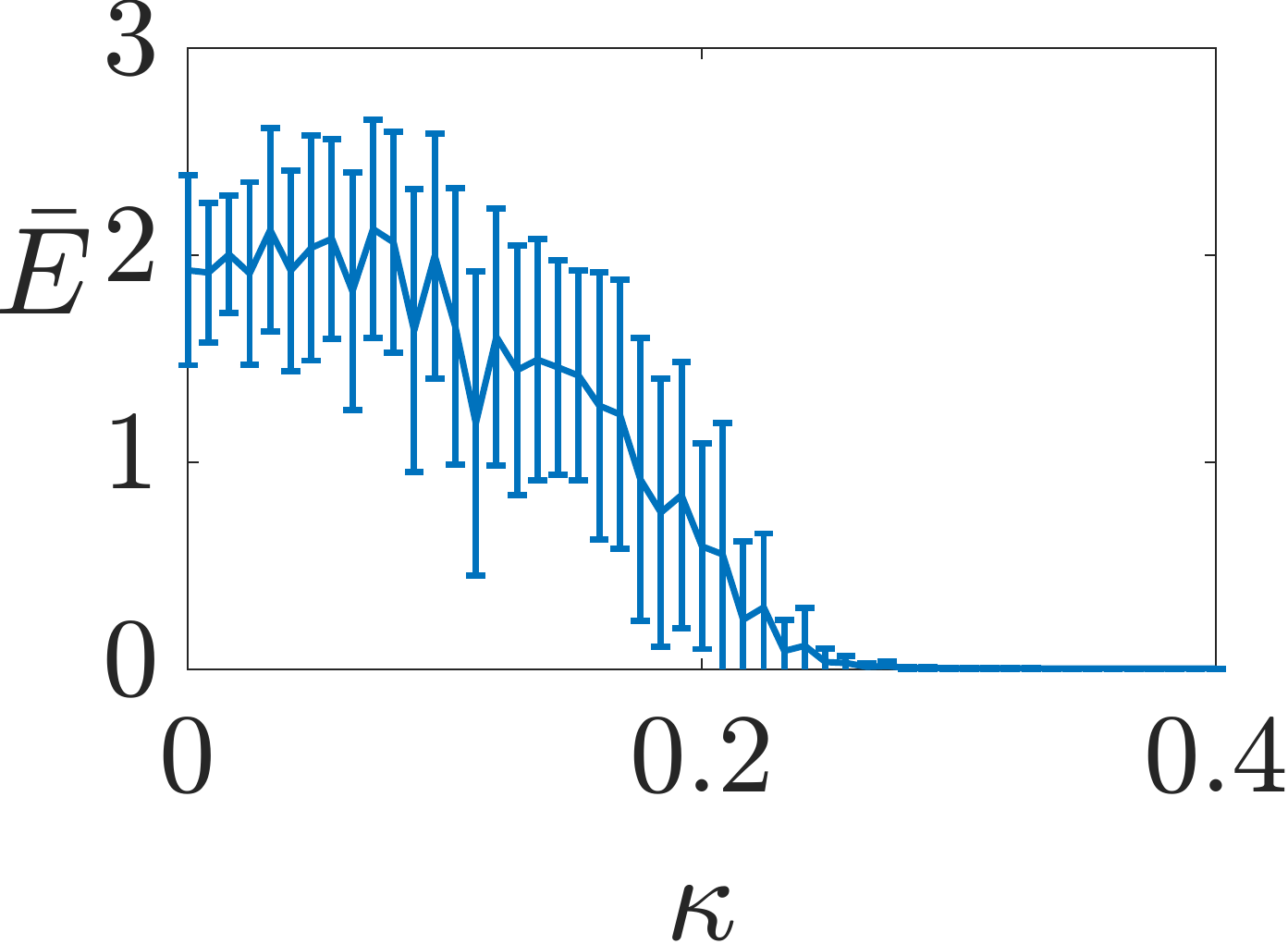} \\  
    \hspace{0.25\linewidth} \text{(C)} \\
    \hspace{0.25\linewidth} \includegraphics[width=.45\linewidth]{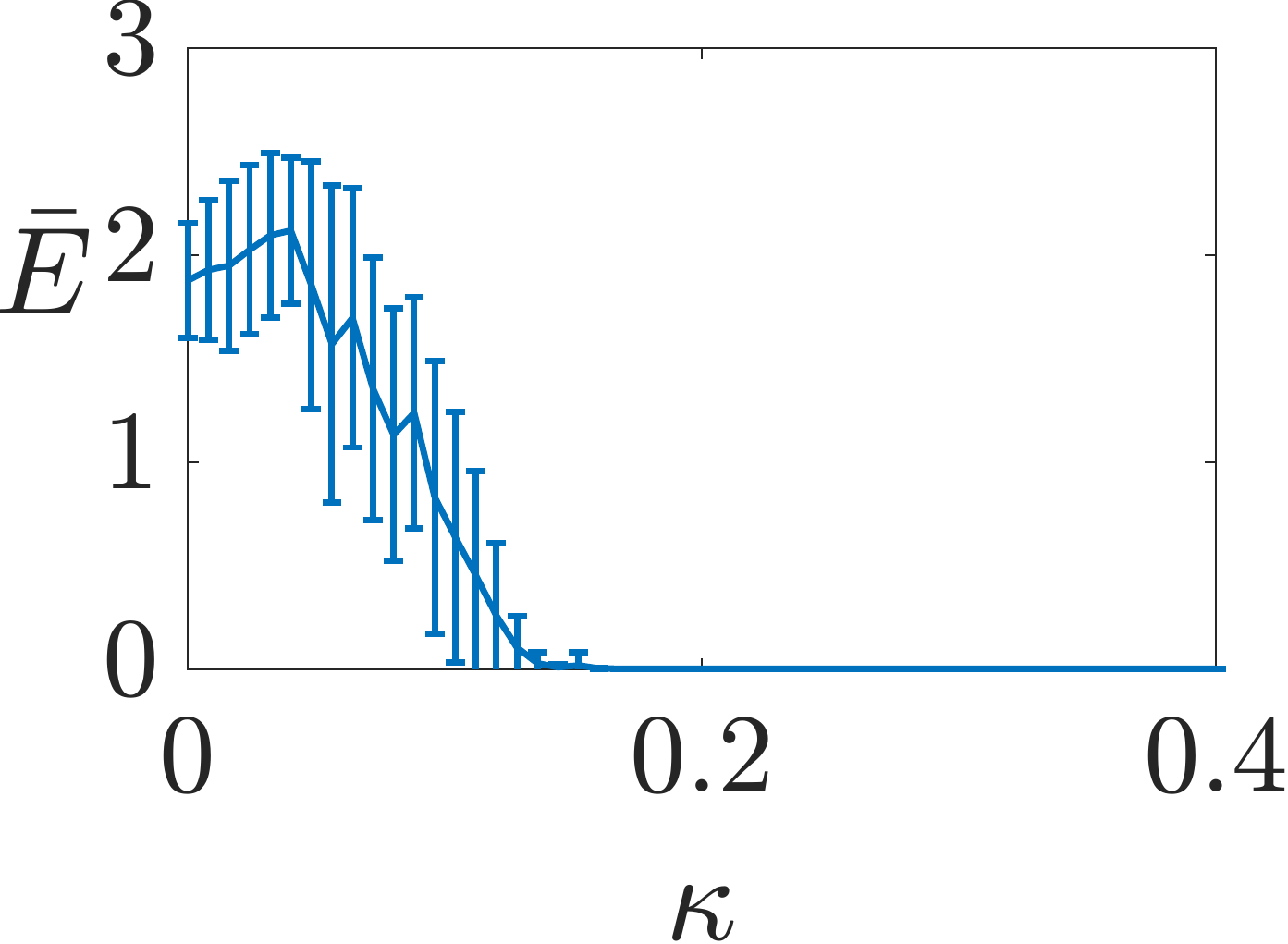}
    \end{tabular}
    \caption{R{\"o}ssler system. (A) Coupling $x_D \rightarrow x_R$ (B) Coupling $\hat{y}_D \rightarrow y_R$ (C) Coupling is both $x_D \rightarrow x_R$ and $\hat{y}_D \rightarrow y_R$. The errorbars represent the standard deviations.  } \label{fig:rossler}
\end{figure}


Figure \ref{fig:rossler} shows a clear advantage of incorporating the RC into the control loop. In this figure, we set $\alpha=0.002$ and $\rho=0.9$. In plot (A) we show the synchronization error versus the coupling strength for the accessible coupling, $x_D \rightarrow x_R$. In plot (B) we show that $\hat{y}_D$, generated by the RC, is sufficient enough to drive the synchronization error close to zero. Plot (C) shows that when both couplings are used, a lower value for the coupling strength is needed to achieve synchronization.

\subsection{Unstable periodic orbits}

Reservoir computing has been previously used to detect unstable periodic orbits (UPOs) of chaotic systems \cite{zhu2019detecting}.
In this section, we aim to synchronize a response system on the trajectory of a drive system, evolving on a UPO. The time-evolutions of the drive and the response systems are still described by Eq.\ \eqref{eq:general}, but 
we consider the special case that the drive system is on a UPO embedded in the chaotic attractor.

 Without loss of generality, we focus on the case of the R{\"o}ssler system, obeying Eq.\,\eqref{eq:rossler}, with the same parameters used before, $a=b=0.2$ and $c=9$. 
First, we find the trajectory of an unstable periodic orbit of the R{\"o}ssler system 
by
using the MatCont toolbox \cite{dhooge2003matcont} for MATLAB.
We used the period-1 trajectory as the drive signal, as this is expected to be the most difficult to synchronize \cite{hunt1996optimal}. 
The reservoir is trained on the UPO signal using 100 nodes, $\alpha = 9.7(10^{-4}), \rho = 0.9$, and input weights $\bw_{\text{in}}$ randomly drawn from a uniform distribution between $-0.5$ and $0.5$; the reservoir then produces an estimate of the unavailable $y$ state from knowledge of the available $x$ state of the UPO signal.
\begin{figure}
    \centering
    \begin{tabular}{l c}
    \text{(A)} \\   
    \includegraphics[width=0.9\linewidth]{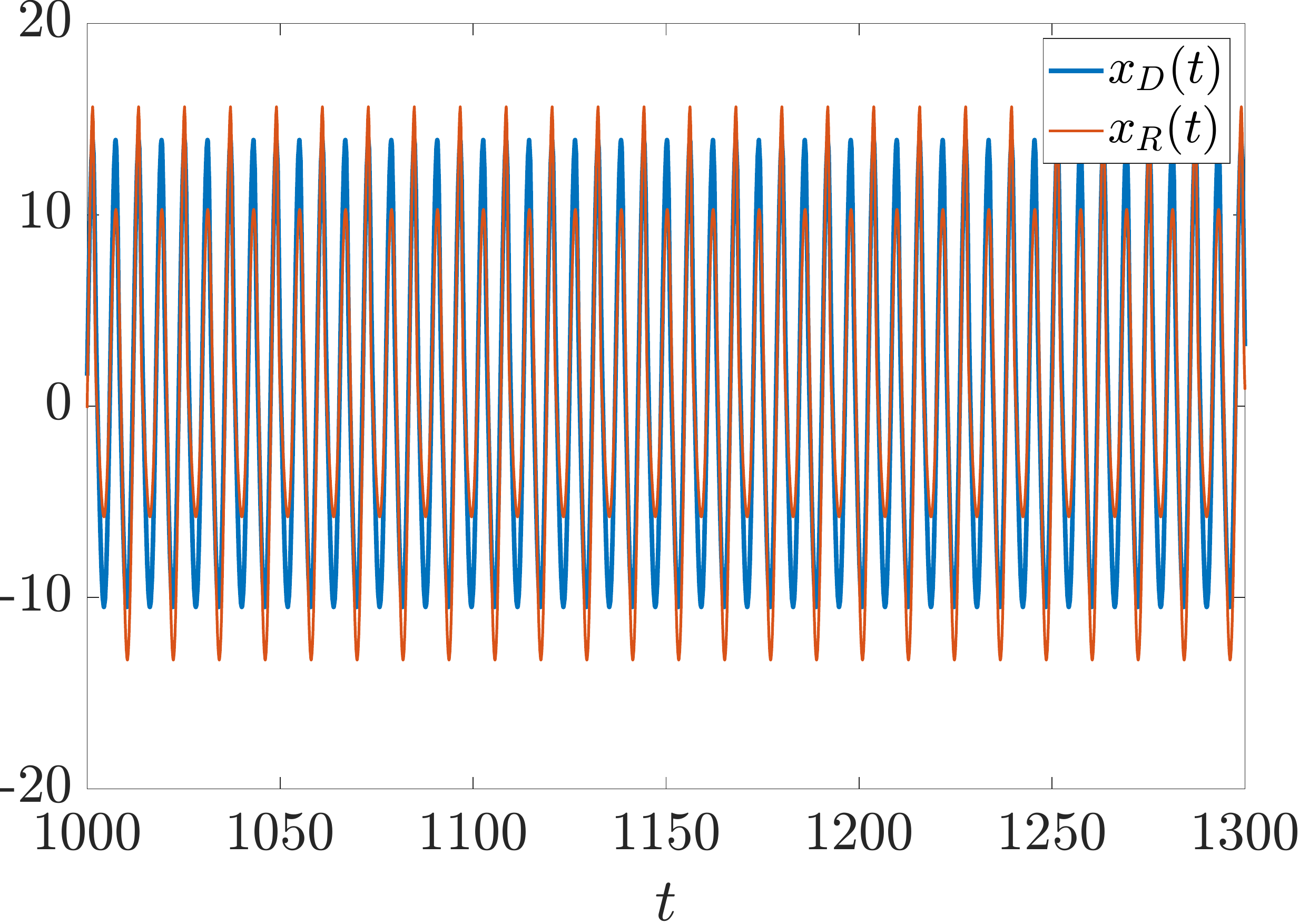} \\
    \text{(B)}\\
    \includegraphics[width=.9\linewidth]{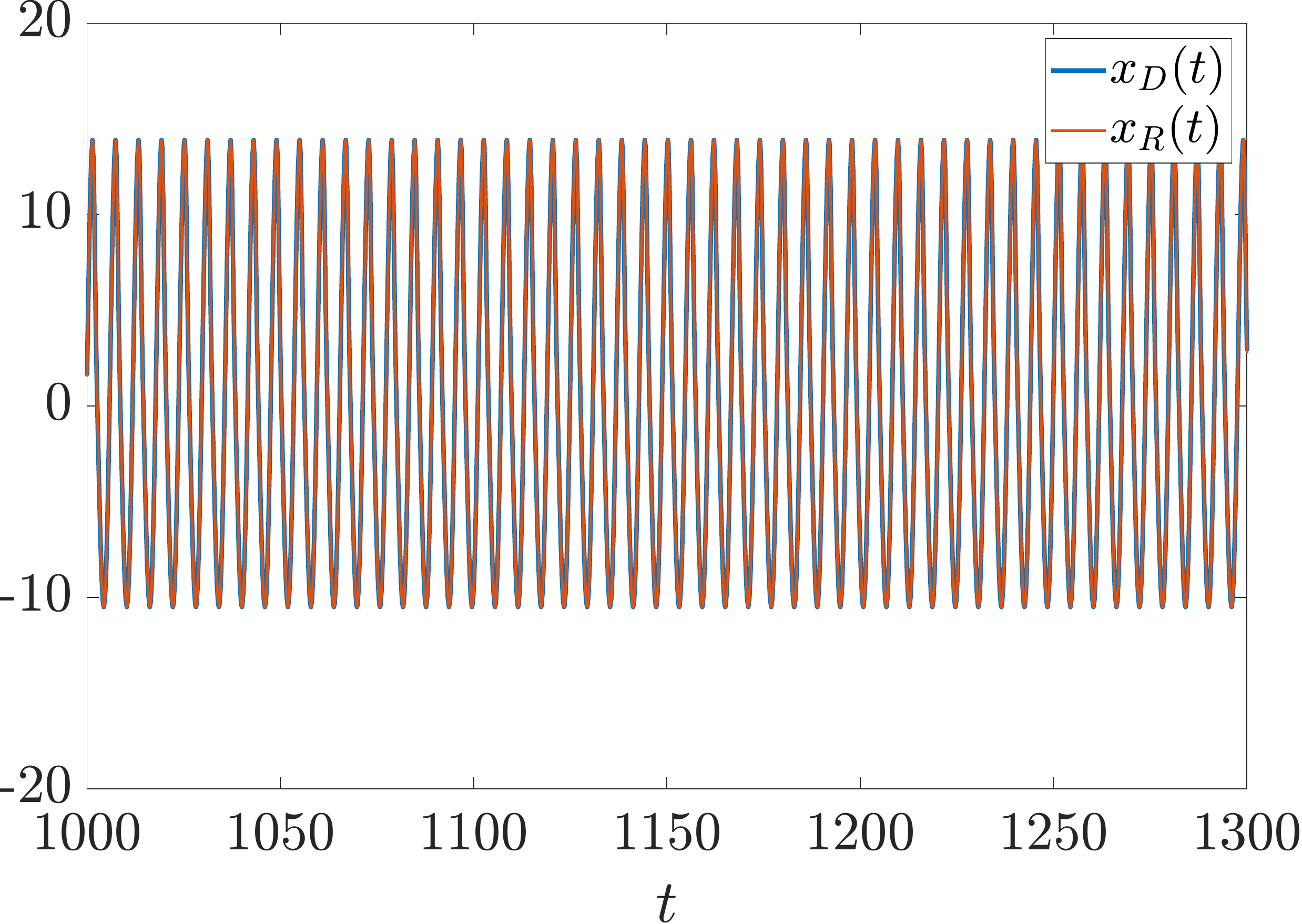}
    \end{tabular}
    \caption{Time trajectories of the $x$ components of the drive and the response systems when the drive R{\"o}ssler system evolves on an unstable periodic orbit. In (A), the two systems are coupled with $x_D \rightarrow x_R$ with the coupling strength $\kappa = 0.3$. In (B), the two systems are coupled with $x_D \rightarrow x_R$ and $\hat{y}_D \rightarrow y_R$ with the same $\kappa = 0.3$. The estimation $\hat{y}_D$ is done through a reservoir computer, as explained in the text.}
    \label{fig:upo}
\end{figure}

Figure \ref{fig:upo} shows that by using the RC we are successful in  achieving synchronization on the period-1 UPO  with a coupling strength ($\kappa=0.3$) for which the drive-response system could not synchronize only through coupling in the $x$ component. 

\color{black}
\subsection{Noise} \label{sec:noise}

In order to compare the results of the RC fairly with ideal knowledge, we consider the situation in which there may be measurement noise present in data acquisition. We proceed under the assumption that all the information obtained from the drive is noise corrupted. This holds true both during the training of the RC as well as implementing it in the control configuration. Following Ref \cite{10.1063/5.0130278}, we set the magnitude of noise present in the training and control configurations to be equal. Our noise corrupted signals are then,
\begin{equation} \label{eq:noise}
    \tilde{\bx}_D(t) = \bx_D(t) + \epsilon \sqrt{\frac{\Delta t}{T}} \pmb{\zeta}(t), 
\end{equation} 
$\tilde{\bx}_D(t)=[\tilde{x}_D(t),\tilde{y}_D(t),\tilde{z}_D(t)]$,
where $\epsilon$ is the magnitude of noise, $T$ is the approximate period of oscillation, and $\pmb{\zeta}(t)$ is a vector with the same dimension of $\bx_D(t)$ composed of elements randomly drawn from a standard normal distribution. 

\begin{figure}
    \centering
    \begin{tabular}{c}
    \text{(A)} \\  
    \includegraphics[width=.8\linewidth]{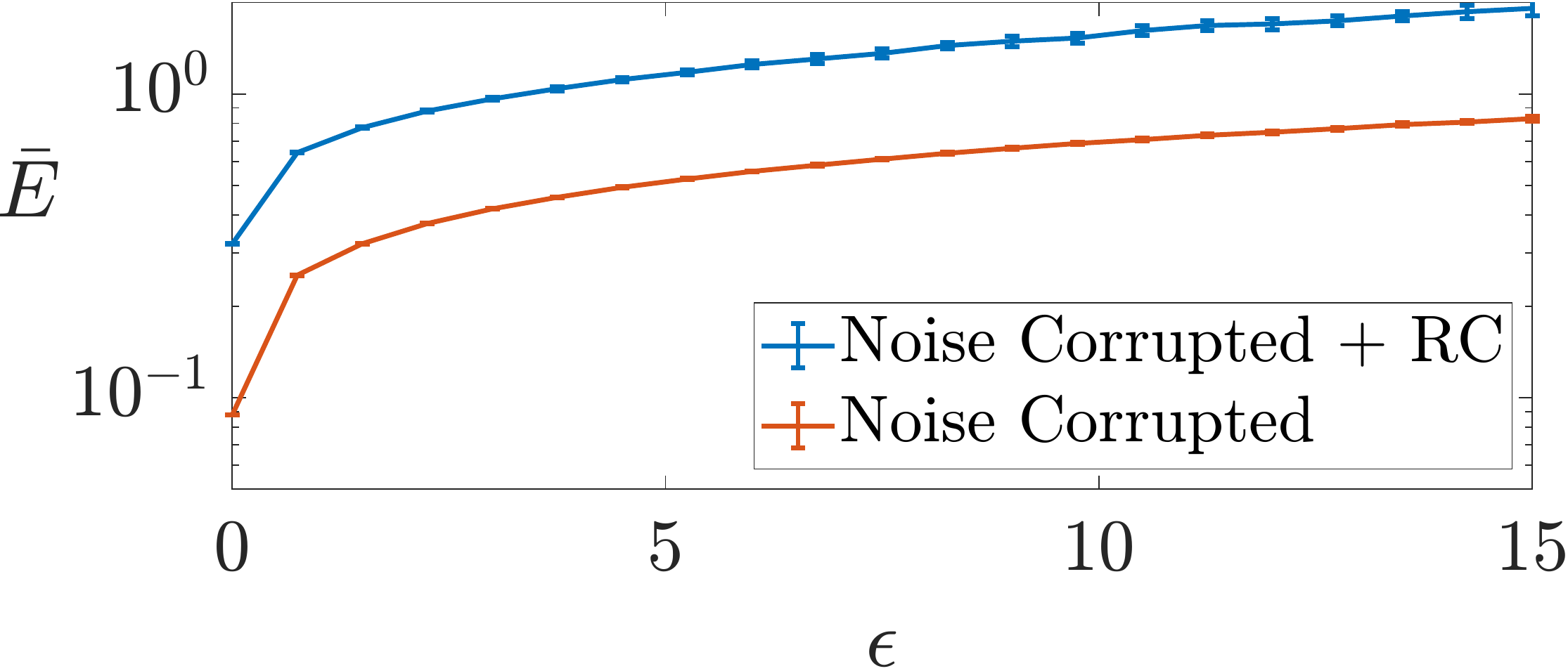} \\
    \text{(B)}\\
    \includegraphics[width=.8\linewidth]{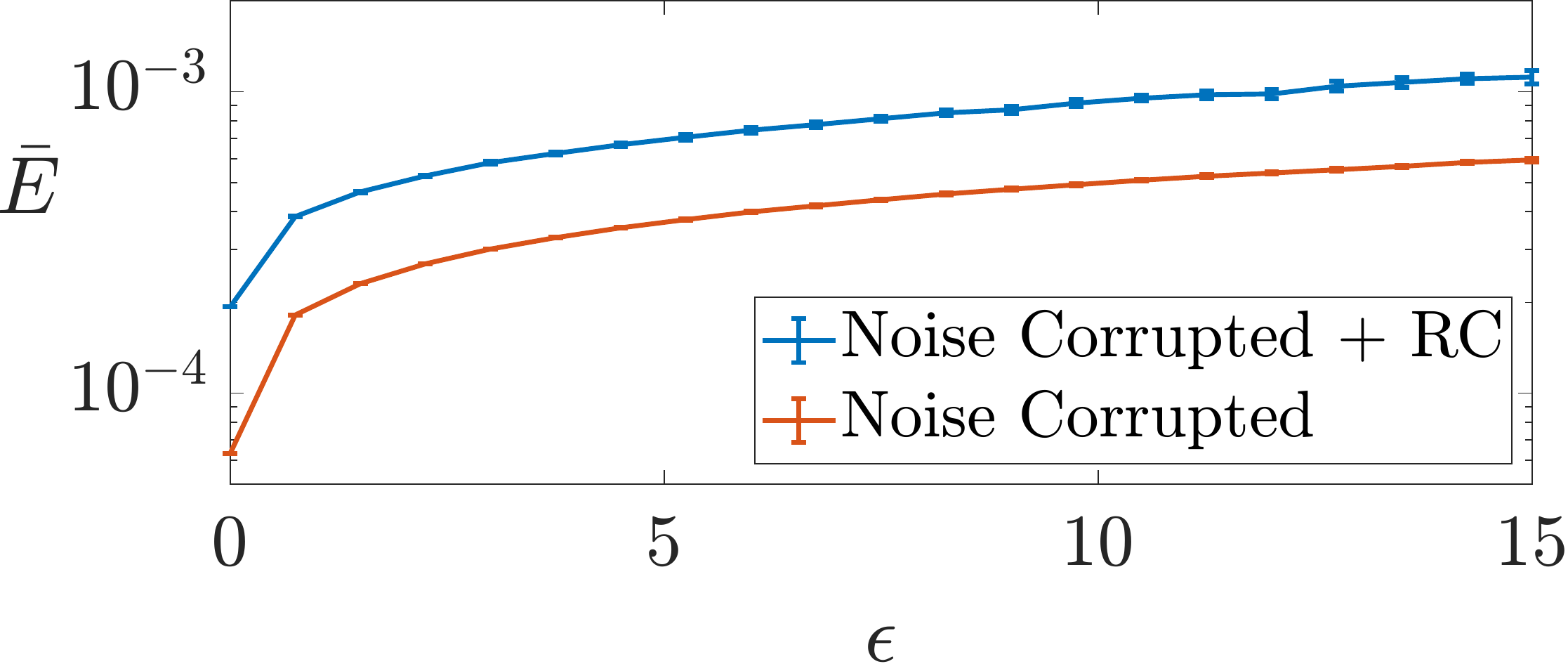}
    \end{tabular}
    \caption{R{\"o}ssler system. We plot the synchronization error as a function of the magnitude of noise, $\epsilon$, as seen in Eq. \eqref{eq:noise}. (A) compares
the case of coupling $\tilde{y}_D \rightarrow y_R$ (noise corrupted) with the case  $\hat{y}_D \rightarrow y_R$ (noise corrupted+RC). (B) compares the case of coupling $\tilde{y}_D \rightarrow y_R$ and $\tilde{x}_D \rightarrow x_R$ (noise corrupted) with the case  $\hat{y}_D \rightarrow y_R$ and $\tilde{x}_D \rightarrow x_R$  (noise corrupted+RC). } \label{fig:rosslernoise}
\end{figure}

We consider the same case of the R\"ossler system described in Sec.\ \ref{sec:ross} and  adopt the same trained RC used in Fig.\ \ref{fig:rossler} with coupling strength $\kappa = 0.2$. In Fig.\ \ref{fig:rosslernoise} we plot $\bar{E}$ as we vary the magnitude of the noise $\epsilon$.  In (A) we compare
the case of coupling $\tilde{y}_D \rightarrow y_R$ (noise corrupted) with the case  $\hat{y}_D \rightarrow y_R$ (noise corrupted+RC). In (B) we compare the case of coupling $\tilde{y}_D \rightarrow y_R$ and $\tilde{x}_D \rightarrow x_R$ (noise corrupted) with the case  $\hat{y}_D \rightarrow y_R$ and $\tilde{x}_D \rightarrow x_R$  (noise corrupted+RC).


We see that for all values of $\epsilon$ in the plotted range, the error is higher in the case in which $\hat{y}_D$ is reconstructed using an RC observer than in the case in which the signal is actually available at the receiver but corrupted with noise. The presence of noise results in a deterioration of the RC performance but overall our proposed strategy based on RC observation appears to be quite robust to the presence of noise, even for large values of $\epsilon$. 
This is true for both the  coupling schemes shown in Fig.\ \ref{fig:rosslernoise}(A) and (B).

\section{Conclusions} \label{sec:conclusions}
In this paper, we used reservoir computing as an observer within a control loop  to estimate the unmeasurable states of a system from its measurable states.  We consider two identical chaotic systems: a drive system undirectionally coupled to a response system.
We have successfully applied this approach to reconstruct unavailable states of the drive system at the response system and to isochronally synchronize the response to the drive. 

In both the cases of the Chen system and of the R\"ossler system, we have shown that usage of the reservoir observer allows us to achieve synchronization for lower values of the coupling strength than it would be possible otherwise. This is particularly relevant to practical situations in which physical constraints may affect our ability to freely choose the strength of the coupling. We showed successful implementation of our proposed scheme to control the state of the response system on an unstable periodic orbit embedded within the chaotic attractor.
The simulations with noisy measured data also demonstrated that the reservoir computer can very well estimate the state trajectory even in the presence of measurement noise.

\appendix

\section{Numerical integration} \label{appendix}
Here, we discuss the numerical method used to integrate the continuous-time dynamics of the drive-response system when they are coupled through the discrete-time reservoir computer.

After the training and testing phases are done, a random point from the testing trajectory is selected to be set as the initial condition for the drive system.
We denote this by $\bx_0 = \bx(t_s)$ where $t_s$ is the randomly chosen time-point from the testing signal.
The corresponding $\br(t_s)$ from the testing phase is set as the initial condition for Eq.\,\eqref{eq:restr}.
A different random point on the training signal is chosen as the initial condition for the response system.
\begin{algorithm}[H]
 \caption{The modified 4th order Runge-Kutta}
 \begin{algorithmic}[1] 
 \Require $\tilde{\bF}(t, \by, \hat{y}_D), \by(t_0), \br(t_0), t, h, K, \alpha, A,  \bw_{\text{in}}, \bw_{\text{out}}$
 \Ensure  $\by, \br$
 \State $k \leftarrow K$
 \For{$j = 0, \hdots, n-1$}
    \State $\by_j \leftarrow \by(t_j)$
    \State $\hat{y}_D \leftarrow \br(t_j)^\top \bw_{\text{out}}$
    \State $k_1 \leftarrow \tilde{\bF}(t_j, \by_j, \hat{y}_D)$
    \State $k_2 \leftarrow \tilde{\bF}(t_j + \frac{h}{2}, \by_j + \frac{h}{2}k_1, \hat{y}_D)$
    \State $k_3 \leftarrow \tilde{\bF}(t_j + \frac{h}{2}, \by_j + \frac{h}{2}k_2, \hat{y}_D)$
    \State $k_4 \leftarrow \tilde{\bF}(t_j + h, \by_j + h k_1, \hat{y}_D)$
    \State $\by(t_{j+1}) \leftarrow \by_j + \frac{h}{6}(k_1 + 2k_2 + 2k_3 + k_4)$
    \If{$k = K$}
        \State  \begin{varwidth}[t]{0.85\linewidth} $\br(t_{j+1}) \leftarrow (1-\alpha)\br(t_j) +  \alpha \tanh (A\br(t_j) + \allowbreak {\qquad \qquad x_D}(t_j)\bw_{\text{in}})$ \end{varwidth}
        \State $k \leftarrow 1$
    \Else
        \State $\br(t_{j+1}) \leftarrow \br(t_j)$
        \State $k \leftarrow k + 1$
    \EndIf
\EndFor \\
\Return $\by, \br$
 \end{algorithmic}  \label{algo}
 \end{algorithm}
The integration of Eq.\,\eqref{eq:general} is done using the 4th-order fixed step Runge-Kutta method.
We assume the $x_D(t)$ component of the drive signal is known, and we try to estimate $\hat{y}_D(t)$. 
We use the compact notation for Eq.\,\eqref{eq:general} as $\by(t) = \Tilde{\bF} (t, \by(t), \hat{y}_D(t))$.
The integration interval is $0 \leq t \leq T_f$ with time points $t_j$, $j = 0, \hdots, n$ sampled at a fixed sampling time $h$. 
For simplicity, we assume  $\Delta t$, the time step of the RC in Eq.\,\eqref{eq:restr}, to be an integer multiple of $h$, i.e., $K = \Delta t / h$ where $K \geq 1$ is an integer.
For large enough $K$, the integration procedure using the Runge-Kutta method has an acceptable numerical error.
For our simulations in this paper, we set $K = 1$, since $h = \Delta t = 0.001$ showed to provide good accuracy for the integration process.
See Algorithm \ref{algo} for the pseudo-code of the modified 4th order Runge-Kutta for the case that $x_D(t)$ of the drive system is known and the goal is to estimate $\hat{y}_D (t)$ of the drive system.
\color{black}



\section*{Acknowledgement} This work was partly funded by NIH Grant No. 1R21EB028489-01A1 and by the Naval Research Lab’s Basic Research Program.
\section*{Data Availability}
The data that support the findings of this study are available
within the article.

 \newcommand{\noop}[1]{}


\begin{thebibliography}{10}

\bibitem{aranson1989nontrivial}
I.~Aranson and N.~Rul'kov, ``Nontrivial structure of synchronization zones in
  multidimensional systems,'' {\em Physics Letters A}, vol.~139, no.~8,
  pp.~375--378, 1989.

\bibitem{pikovskii1984synchronization}
A.~S. Pikovskii, ``Synchronization and stochastization of array of self-excited
  oscillators by external noise,'' {\em Radiophysics and Quantum Electronics},
  vol.~27, no.~5, pp.~390--395, 1984.

\bibitem{FujiYama83}
H.~Fujisaka and T.~Yamada, ``Stability theory of synchronized motion in
  coupled-oscillator systems,'' {\em Prog. Theor. Phys.}, vol.~69, no.~1,
  pp.~32--47, 1983.

\bibitem{afraimovich1986stochastic}
V.~Afraimovich, N.~Verichev, and M.~Rabinovich, ``Stochastic synchronization of
  oscillations in dissipative systems,'' {\em Radiofizika}, vol.~29, no.~9,
  pp.~1050--1060, 1986.

\bibitem{pecora1990synchronization}
L.~M. Pecora and T.~L. Carroll, ``Synchronization in chaotic systems,'' {\em
  Physical review letters}, vol.~64, no.~8, p.~821, 1990.

\bibitem{pecora1998master}
L.~M. Pecora and T.~L. Carroll, ``Master stability functions for synchronized
  coupled systems,'' {\em Physical review letters}, vol.~80, no.~10, p.~2109,
  1998.

\bibitem{nazerian2022matryoshka}
A.~Nazerian, S.~Panahi, I.~Leifer, D.~Phillips, H.~A. Makse, and F.~Sorrentino,
  ``Matryoshka and disjoint cluster synchronization of networks,'' {\em Chaos:
  An Interdisciplinary Journal of Nonlinear Science}, vol.~32, no.~4,
  p.~041101, 2022.

\bibitem{pecora2015synchronization}
L.~M. Pecora and T.~L. Carroll, ``Synchronization of chaotic systems,'' {\em
  Chaos: An Interdisciplinary Journal of Nonlinear Science}, vol.~25, no.~9,
  p.~097611, 2015.

\bibitem{iglesias2010control}
P.~A. Iglesias and B.~P. Ingalls, {\em Control theory and systems biology}.
\newblock MIT press, 2010.

\bibitem{del2016control}
D.~Del~Vecchio, A.~J. Dy, and Y.~Qian, ``Control theory meets synthetic
  biology,'' {\em Journal of The Royal Society Interface}, vol.~13, no.~120,
  p.~20160380, 2016.

\bibitem{liu2016control}
Y.-Y. Liu and A.-L. Barab{\'a}si, ``Control principles of complex systems,''
  {\em Reviews of Modern Physics}, vol.~88, no.~3, p.~035006, 2016.

\bibitem{klickstein2017energy}
I.~Klickstein, A.~Shirin, and F.~Sorrentino, ``Energy scaling of targeted
  optimal control of complex networks,'' {\em Nature Communications}, vol.~8,
  2017.

\bibitem{jaeger2001echo}
H.~Jaeger, ``The “echo state” approach to analysing and training recurrent
  neural networks-with an erratum note,'' {\em Bonn, Germany: German National
  Research Center for Information Technology GMD Technical Report}, vol.~148,
  no.~34, p.~13, 2001.

\bibitem{maass2002real}
W.~Maass, T.~Natschl{\"a}ger, and H.~Markram, ``Real-time computing without
  stable states: A new framework for neural computation based on
  perturbations,'' {\em Neural computation}, vol.~14, no.~11, pp.~2531--2560,
  2002.

\bibitem{pathak2017using}
J.~Pathak, Z.~Lu, B.~R. Hunt, M.~Girvan, and E.~Ott, ``Using machine learning
  to replicate chaotic attractors and calculate lyapunov exponents from data,''
  {\em Chaos: An Interdisciplinary Journal of Nonlinear Science}, vol.~27,
  no.~12, p.~121102, 2017.

\bibitem{lu2017reservoir}
Z.~Lu, J.~Pathak, B.~Hunt, M.~Girvan, R.~Brockett, and E.~Ott, ``Reservoir
  observers: Model-free inference of unmeasured variables in chaotic systems,''
  {\em Chaos: An Interdisciplinary Journal of Nonlinear Science}, vol.~27,
  no.~4, p.~041102, 2017.

\bibitem{carroll2022time}
T.~L. Carroll and J.~D. Hart, ``Time shifts to reduce the size of reservoir
  computers,'' {\em Chaos: An Interdisciplinary Journal of Nonlinear Science},
  vol.~32, no.~8, p.~083122, 2022.

\bibitem{hart2023time}
J.~D. Hart, F.~Sorrentino, and T.~L. Carroll, ``Time-shift selection for
  reservoir computing using a rank-revealing qr algorithm,'' {\em Chaos: An
  Interdisciplinary Journal of Nonlinear Science}, vol.~33, no.~4, 2023.

\bibitem{cunillera2019cross}
A.~Cunillera, M.~C. Soriano, and I.~Fischer, ``Cross-predicting the dynamics of
  an optically injected single-mode semiconductor laser using reservoir
  computing,'' {\em Chaos: An Interdisciplinary Journal of Nonlinear Science},
  vol.~29, no.~11, p.~113113, 2019.

\bibitem{kong2023reservoir}
L.-W. Kong, Y.~Weng, B.~Glaz, M.~Haile, and Y.-C. Lai, ``Reservoir computing as
  digital twins for nonlinear dynamical systems,'' {\em Chaos: An
  Interdisciplinary Journal of Nonlinear Science}, vol.~33, no.~3, 2023.

\bibitem{zimmermann2018observing}
R.~S. Zimmermann and U.~Parlitz, ``Observing spatio-temporal dynamics of
  excitable media using reservoir computing,'' {\em Chaos: An Interdisciplinary
  Journal of Nonlinear Science}, vol.~28, no.~4, p.~043118, 2018.

\bibitem{SYNCBOOK}
S.~Strogatz, {\em Sync: The Emerging Science of Spontaneous Order}.
\newblock Hyperion. New York, 2003.

\bibitem{antonik2018using}
P.~Antonik, M.~Gulina, J.~Pauwels, and S.~Massar, ``Using a reservoir computer
  to learn chaotic attractors, with applications to chaos synchronization and
  cryptography,'' {\em Physical Review E}, vol.~98, no.~1, p.~012215, 2018.

\bibitem{weng2019synchronization}
T.~Weng, H.~Yang, C.~Gu, J.~Zhang, and M.~Small, ``Synchronization of chaotic
  systems and their machine-learning models,'' {\em Physical Review E},
  vol.~99, no.~4, p.~042203, 2019.

\bibitem{hu2022synchronization}
W.~Hu, Y.~Zhang, R.~Ma, Q.~Dai, and J.~Yang, ``Synchronization between two
  linearly coupled reservoir computers,'' {\em Chaos, Solitons \& Fractals},
  vol.~157, p.~111882, 2022.

\bibitem{weng2023synchronization}
T.~Weng, X.~Chen, Z.~Ren, H.~Yang, J.~Zhang, and M.~Small, ``Synchronization of
  machine learning oscillators in complex networks,'' {\em Information
  Sciences}, vol.~630, pp.~74--81, 2023.

\bibitem{hart2023estimating}
J.~D. Hart, ``Estimating the master stability function from the time series of
  one oscillator via machine learning,'' {\em arXiv preprint arXiv:2304.13125},
  2023.

\bibitem{canaday2021model}
D.~Canaday, A.~Pomerance, and D.~J. Gauthier, ``Model-free control of dynamical
  systems with deep reservoir computing,'' {\em Journal of Physics:
  Complexity}, vol.~2, no.~3, p.~035025, 2021.

\bibitem{zhu2019detecting}
Q.~Zhu, H.~Ma, and W.~Lin, ``Detecting unstable periodic orbits based only on
  time series: When adaptive delayed feedback control meets reservoir
  computing,'' {\em Chaos: An Interdisciplinary Journal of Nonlinear Science},
  vol.~29, no.~9, p.~093125, 2019.

\bibitem{Pe:Ca}
L.~Pecora and T.~Carroll, ``Master stability functions for synchronized coupled
  systems,'' {\em Phys. Rev. Lett.}, vol.~80, pp.~2109--2112, 1998.

\bibitem{Pecora2009}
L.~Huang, Q.~Chen, Y.-C. Lai, and L.~M. Pecora, ``Generic behavior of
  master-stability functions in coupled nonlinear dynamical systems,'' {\em
  Phys. Rev. E}, vol.~80, p.~036204, 2009.

\bibitem{dhooge2003matcont}
A.~Dhooge, W.~Govaerts, and Y.~A. Kuznetsov, ``Matcont: A matlab package for
  numerical bifurcation analysis of odes,'' {\em ACM Trans. Math. Softw.},
  vol.~29, p.~141–164, jun 2003.

\bibitem{hunt1996optimal}
B.~R. Hunt and E.~Ott, ``Optimal periodic orbits of chaotic systems,'' {\em
  Physical review letters}, vol.~76, no.~13, p.~2254, 1996.

\bibitem{10.1063/5.0130278}
C.~Nathe, C.~Pappu, N.~A. Mecholsky, J.~Hart, T.~Carroll, and F.~Sorrentino,
  ``{Reservoir computing with noise},'' {\em Chaos: An Interdisciplinary
  Journal of Nonlinear Science}, vol.~33, 04 2023.
\newblock 041101.

\end{thebibliography}
\end{document}